 \documentclass[final,5p,times,twocolumn]{elsarticle}

 \usepackage{epsfig}

\usepackage{amssymb}

\journal{Arxiv}

\begin{document}

\begin{frontmatter}



\title{Brownian dynamic simulation by reticular mapping matrix method.}


\author{Eric Plaza}

\address{Instituto Zuliano de Investigaciones Tecnol\'{o}gicas,\\ La Cañada de Urdaneta, Km 15 Via Palmarejo Viejo. \\ Telf 0058-261 7913769, 0058-416 8349193}

\begin{abstract}

This work proposes a method for the two-dimensional simulation of Brownian particles in a fluid with restrictions. The method is based on simple numerical rules between two matrices. One of the matrix represent the identification of all particles over which are adapted  statistics rules for particles movement, the results are mapped over other matrix which represent the particles positions. The rules for the movement are established by a statistic mechanism allowing assign random or non-random movement direction. The same probably of movement for each direction for each time step is assumed, in order to be agreed with the physics conditions of Brownian movement in a two dimensional network. The root mean square displacement of all particles was calculated in a large number of simulations, together with the translational velocity of particles in order to compare with theoretical values of diffusion coefficient and the validation of model. Also, the time duration for some simulations vs. the number of particles and concentration was calculated. 

\end{abstract}


\end{frontmatter}


\section{Introduction}
The study of Brownian particles dynamics has applications directed to suspensions in the nanometers and micrometer scale to be used for industrials process, paint development, pastes and ceramics, were the understanding of its dynamic will collaborate in bio-molecular transport and manipulations, chemical processes, erosion, and so on.

The works in Brownian motion has been directed to understanding the macroscopic properties of fluids by the motion of particles embedded in it. Usually the mathematical treatment of fluid is a small volume element. Differential mathematics takes this as very small compared with the volume of the body under 
consideration, but large compared with the distances between the molecules. 
\cite{001}. 

The dynamic behavior of fluids is governed by the differential Navier Stokes equation, however the analytical resolution in complex situations is limited, and various methods have been proposed for its dynamics simulation through the discretization of differential equations (differences finite, finite element). Other methods such as grids simulate the motion of the particles in this fluids, according to rules that can be random or predetermined. The work of Wolfram \cite{002}  defines these models as cellular automata. From cellular automata, the works of Hardy \emph{et al.} \cite{003,004,005} was developed the HPP model, which describes the motion of particles through movement rules in four directions. This proposal was extended in the work of Frisch et al \cite{006} and Lattice Boltzmann model or FHP, proposing an extension of the model conditions of HPP, considering mass and momentum conservation in particles collisions, and achieving a performance that recreates the Navier-Stokes equations with certain simplifications. Using this model, the work of Nie \emph{et al.} \cite{007} represents the Brownian motion, obtaining its validity by calculating average values of the displacement particles and the self-diffusion coefficient. 

Other fluid particles simulation technique is the molecular dynamics method, which solves the equations of motion for each particle in order to represent the movement of a large set of particles in a fluid medium, validating it in the thermodynamic limit \cite{008}. In the case of Brownian motion the more commonly methodology used for the simulations of this phenomena are represented in the work of Ermak and McCammon \cite{009},  based their numerical method in the Langevin equation. Other proposed methodologies for the particle motion is the SPH (Smoothed Particle Hydrodynamics) \cite{010} \cite{011} that interpolates the fluid properties over the space and particles using a kernel functions, in this model each particle is moved by the kernel which assign directions under statistical restrictions in accordance with physicals rules.

The problem in the simulation methods has been directed towards the development of computational schemes that represent the complexities of physical systems using less computational cost; these conditions require simplify the algorithms in order to save computational resources and improve the efficiencies of programs, some works as Phillips, \emph{et al.} \cite{012} and Hellander \emph{et al.} \cite{013}  show this points improving and reformulating codes in order to optimize the simulations. 

We present an initial work inspired by the simplicity that provide interactions with integers, and handled by a statistical framework, that represent the conditions of Brownian particles motion in a fluid. The algorithm uses a matrix called $"Mi"$ that represent the operations over all particles in the simulation, and another called $"Me"$ representing the positions of these particles in a square grid. The motion of these particles occurs by adding or subtracting random integers to the matrix $Mi$, and mapping these changes to the positions in the Me matrix. The method is called "Reticular Mapping Matrix" (RMM) by the conditioning of mapping rules between the matrix that generate the movement of particles and the matrix which represent the space were the particles are moved. Using these rules we adapt the problem in the terms of Brownian motion particles in adiabatic fluid, two-dimensional restrictions and closed borders. To validate the numerical model, we compute and comparing with theoretical results that including the measurement of the square displacement average of particles in the medium, the average particle velocities, the measure of self-diffusion coefficient and the time durations of simulations in order to evaluate the algorithm capabilities.

\section{Simulation technique }
Brownian motion is the random movement observed in microscopic particles immersed in a fluid; this is caused because the particles surface is bombarded constantly by the fluid molecules subject to thermal agitation. The bombing at the atomic scale is not completely uniform and undergoes pressure variations over the particles surfaces, causing the motion of particles in a random direction.
The mathematical description of the phenomena was developed by the work of Einstein \cite{014} , relating the macroscopic diffusion coefficients D and microscopic properties of matter. For this, the self-diffusion coefficient of particles in a fluid is defined as:

\begin{equation}
D=RT/Na6\eta\pi
\end{equation}

Were "R" is the ideal gas constant, $"n"$ the number of particles, $"T"$ the temperature in Kelvin, $\eta$ viscosity, $"a"$ the radius of the Brownian particle, and $"t"$ time. For Brownian motion Einstein found that the mean displacement of particles $\lambda$ in one direction is an expression such as:

\begin{equation}
\lambda=\sqrt(2Dt)
\end{equation}

The work of Langevin \cite{015} presents the equations governing the motion of these particles, and more recently the work of Chavanis et al. \cite{016} presents a discussion on the validity of the theoretical results. In general, the results obtained in these systems are characteristic relationships of "random walkers" in which the mean square displacement of particles is a linear function of time after a large number of random steps \cite{017}.

This work presents a method for simulating the self-diffusive behavior of interacting particles in a solution through a mapping technique based on rules of movement between two matrices. The theory of Brownian motion was developed in order to describe the dynamic behavior of particles whose mass and size are much larger than the rest of the environment in which they are. In the simulation, each particle is located randomly in a two-dimensional lattice. For each time step a direction of movement has to be assigned randomly. For this, the model used two matrixes, $"Me"$ and $"Mi"$ (see Figure 1). The numbers inside each cell of $"Mi"$ represent the position of each particle in the matrix $Me"$, and the index above the cell represents the "name" or identification of each particle. For the $"Me"$ matrix, the number in the cell represents the identification or "name" of the particles and the index above the positions.  (The dimension of matrix $"Mi"$ will be N1xN2 and N3xN4 for $"Me"$, where the product of $N1.N2$ represents the total number of particles and $N3.N4$ the cell numbers of space).

\begin{figure}
\includegraphics[scale=0.28]{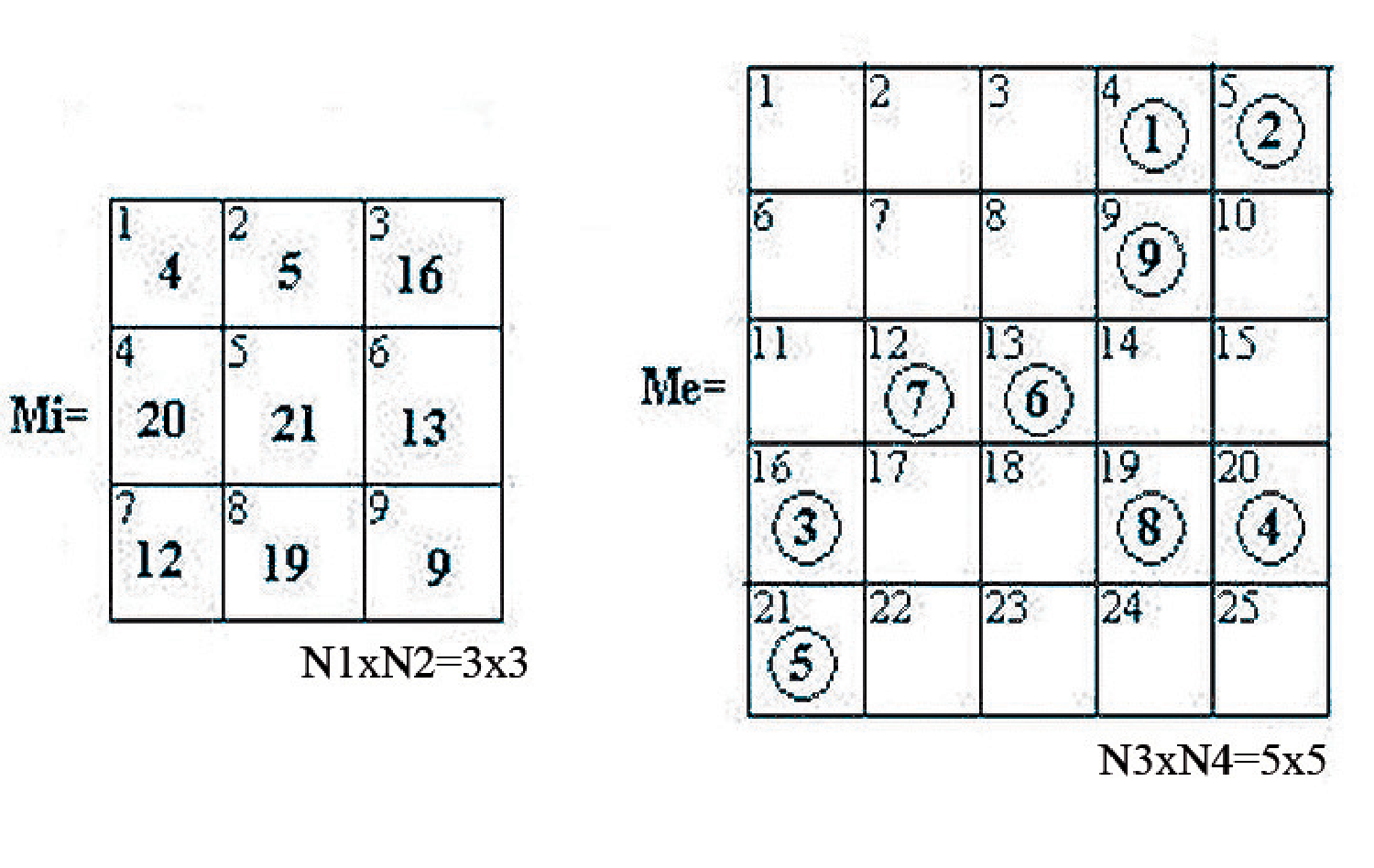}
\caption{Example: The particle 1 of Mi (1x1) is mapped in to cell $"4"$ of $"Me"$ (1x4). The particle 3 of $"Mi"$ is in position 16 of $"Me"$ and so on for all particles.}
\end{figure}

During each time step all particles within the network $"Me"$ will be moved randomly in one of its possible directions. For this two-dimensional case we have five directions of motion (rightward, leftward, upward, downward and zero movement). The motion of each particle is governed by the matrix operations in $"Mi"$, where the randomly sum of integers to each cell of $"Mi"$ determine the direction of motion of the particle in $"Me"$, i.e. the nature of such sum will represent the direction of movement. If the sum is chosen +1 the movement will be directed to right direction, -1 leftward, +N3 movement downward, -N3 upward direction and "0" for movement. This method will be called "Reticular Mapping Matrix" or RMM.

\begin{figure}
\includegraphics[scale=0.44]{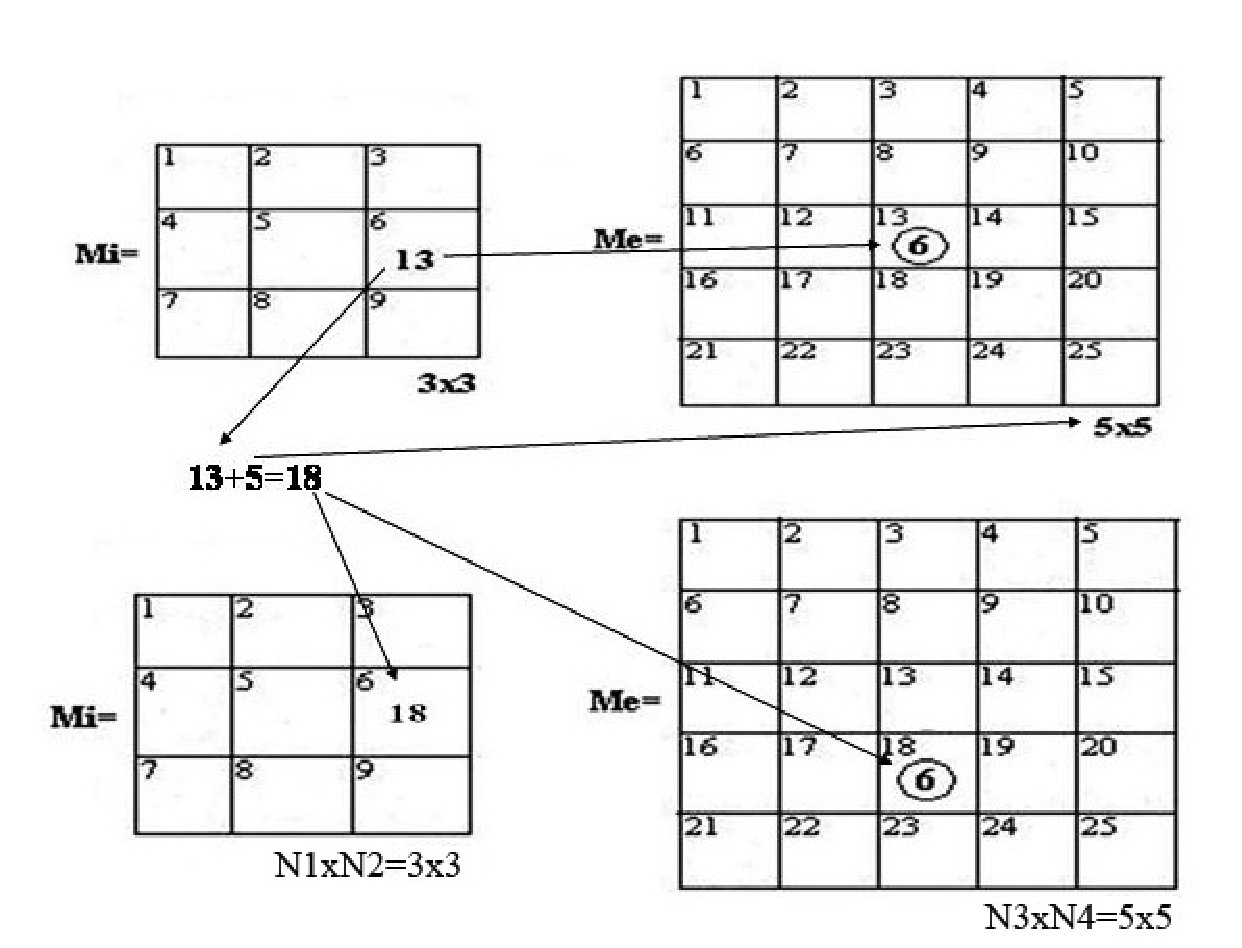}
\caption{Displacement of particle $"6"$, from position $"13"$ to $"18"$. Assuming the operation over the cell (particle 6) of Mi: 13+5 , we perform a movement to the bottom direction (the number 5 correspond to the files number of matrix $"Me"$), and the particle will be moved  to position $"18"$. To move in the left direction we must add (-1), right (+1),  up (-N3), and zero for non movement}
\end{figure}

For each time step, over all particles, is assigned a direction of movement chosen randomly in order to represent the Brownian motion. A mechanism in the program prevents the existence of particles with the same number within the matrix $"Mi"$, preventing particles in the same position, and traffic problems. For the boundary conditions a set rules over the cell borders make a closed boundaries or box like. Additionally, the method allows to easily changing the weight of the direction probabilities, changing the statistical behavior of particles as seen in Figure 3.
 
\begin{figure}
\includegraphics[scale=0.5]{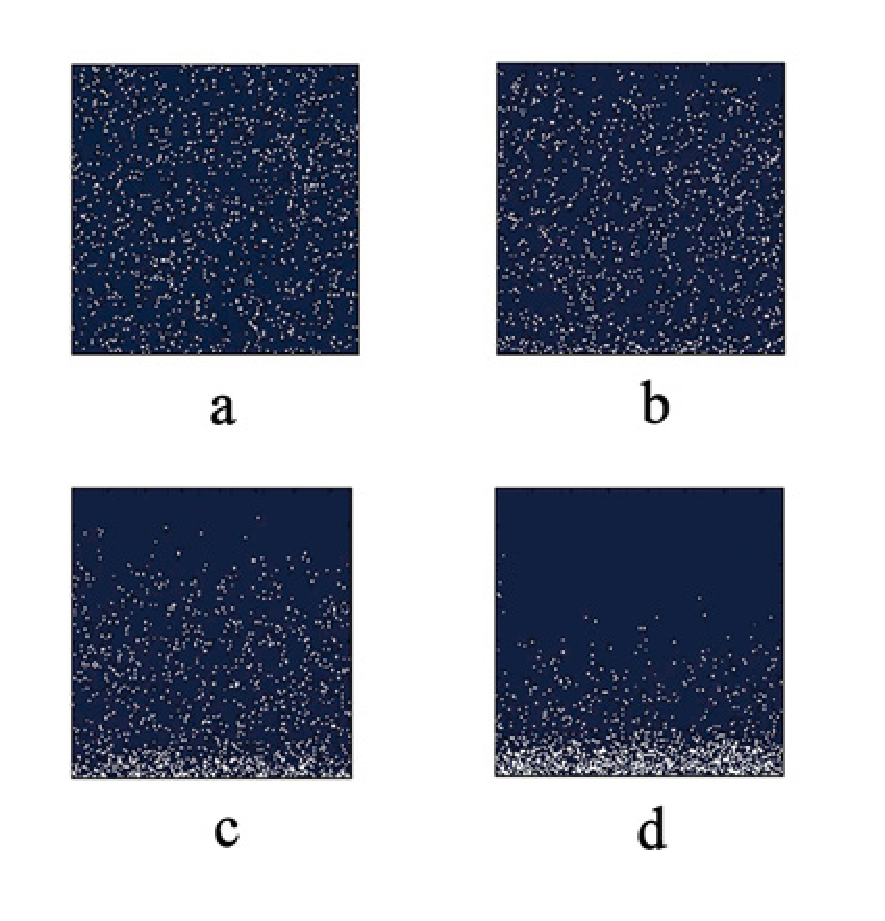}
\caption{Simulation performed for 1000 time steps in a matrix, were $"Me"$=150x150 = 22500 cells, $"Mi"$= 50x50 = 2500 particles. If we use the probability weights: 30\% downward, 20\% upward, 20\% rightward, 20\% leftward and 10\% zero movements, we note that for a considerable number of time steps, the particles tend to go in to bottom simulation area. Figure a: First time step. Figure b: time step number 100. Figure c: time step number 500. Figure d: time step number 1000. }
\end{figure}

In order to define the particles size in simulation according to physical quantities, we assume a square lattice were the particles size is $"a"$ (in this case we take the Einstein's work value  $1x10^{-6}$ m ). We define the size of reference network "d" multiplying the numbers of cells in the row by the size  particles, $d=N3.a$. Using this quantities we can rescale the size of particles changing the number of of cells $"n"$ and fixing the references size of network "d". Figure 4, shows graphically these relationships.

\begin{equation}
a=(d/n)
\end{equation}

In this work we used the particle size values: $a_{1}$ =  $1x10^{-6}$ m (as reference case), $a_{2}$ =  $0.8x10^{-6}$ m, $a_{3}$ = $0.66x10^{-6}$ m, whose networks are formed respectively by, $a_1$: 100x100 (reference case), $a_2$: 125x125 and a3: 150x150.

\begin{figure}
\includegraphics[scale=0.30]{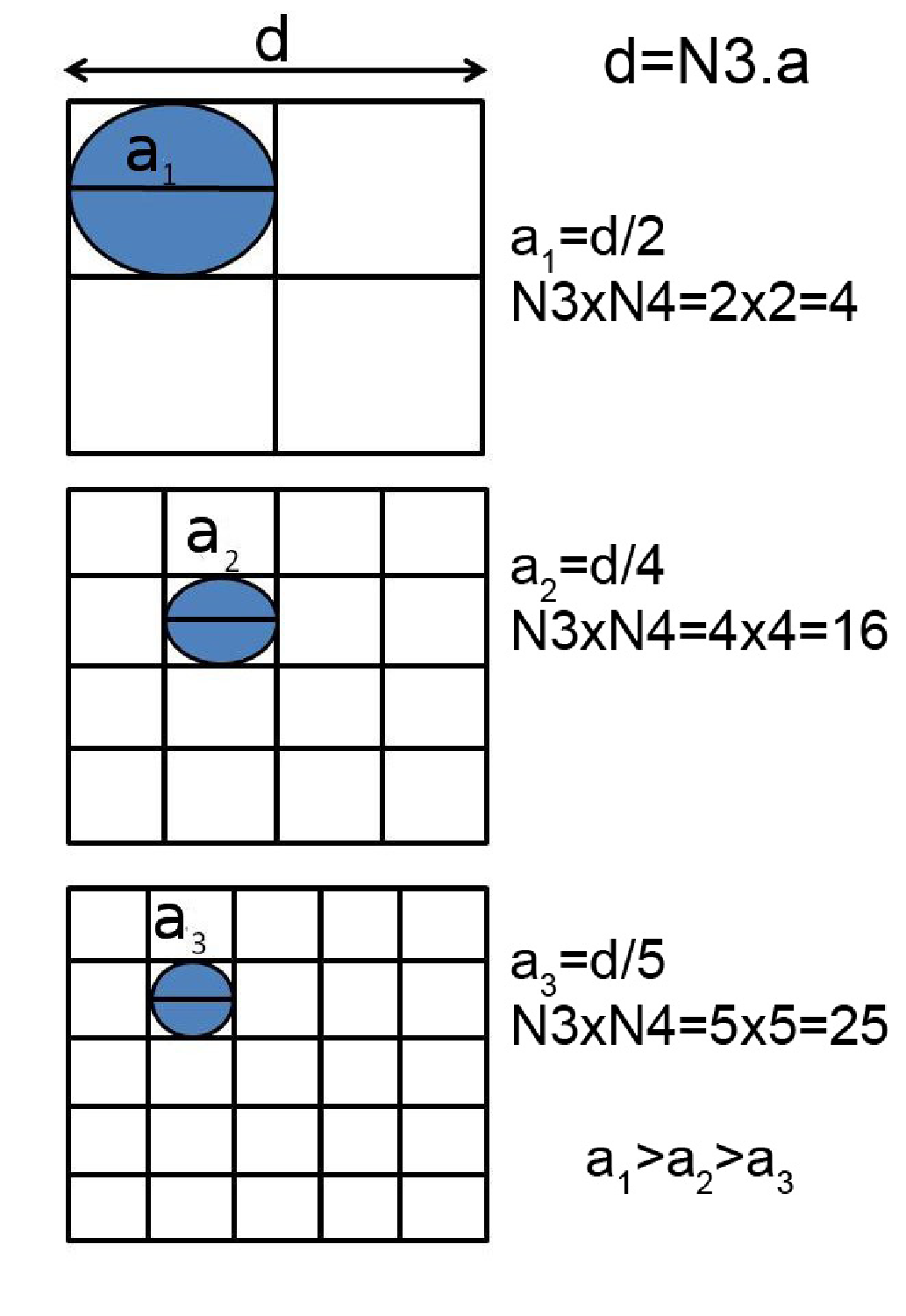}
\caption{Scheme adopted to establish the particle sizes. Setting a reference value for a reference network size $"d"$, is possible to define other particle sizes based on chosen the scale, by changing the number of cells of other networks, keeping the total size of the cell in $"d"$. }
\end{figure}

\section{Discussion and results}

In order to validating results, we compute the mean square displacement of all particles $\lambda$ , associated with the self-diffusion coefficient for simulation $"D_{s}"$ of fluid through the equation (4) \cite{009} \cite{017} where the brackets denote the average of the displacements of all  particles. This expression is a linear function of time after a large number of random steps. To corroborate this condition $\lambda$ was calculated in 500 time steps for 500 simulations.

\begin{equation}
D_{s}=  \frac{  \langle | r_{i}(t)- r_{i}(0)  |^2 \rangle   }{6t}  
\end{equation}

Initial positions are represented by the subscript i and f for the final position. The number of particles is represented by the subscript n, where n = N1xN2. Knowing the initial and final positions of each particle we apply the expression for two-dimensional distance between two points over all particles, for after to computing the average. In order to define the the size of the particles we use the definitions of previous section, were a1 = 1x10-6 m, a2 = 0.8x10-6 m, a3 = 0.66x10-6 m. The results are shown in Figure 5, for 100 simulations, each one with a number of 500 steps, the percentages of weight directions of motion was 20\% right, left, down, up, and zero movement. The volume fraction was set at approximately in 10\% for each case. Figure 5 shows the linear behavior of $\lambda$  for a representative number of simulations as predicted by theory.

The value of $\lambda_{a_{1}}$ is consistent with the theoretical value of Einstein work  $a_{1}=8x10^{-7}$  m when scaling the simulation total time to:  t = 9.8255 s and compute with equation (4). The simulation was carried out in 500 time steps, each time step corresponds to 0.019651 s. The same reasoning is valid to calculate the time in the others simulations.

\begin{figure}
\includegraphics[scale=0.55]{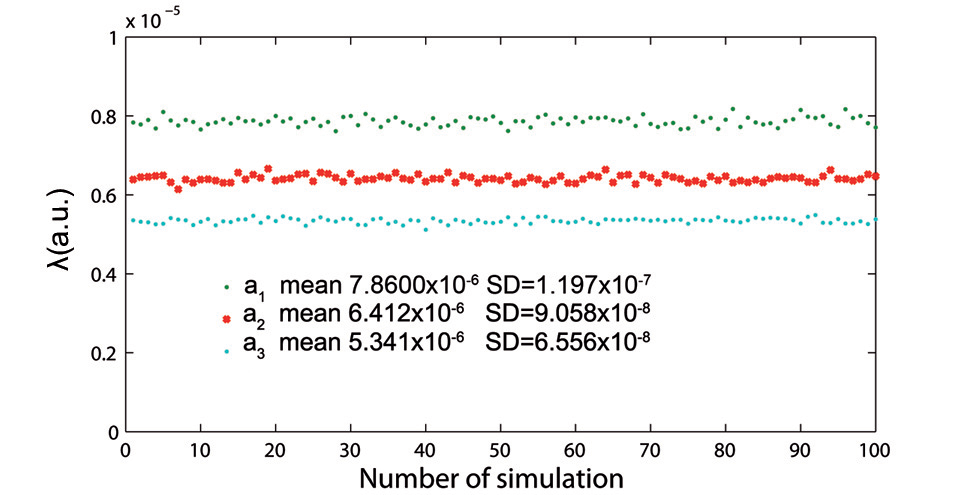}
\caption{ Mean Values and standard deviation (SD) obtained for different particle sizes. $a_{1}=1x10^{-6} m$, $a_{2}=0.8x10^{-6} m$, $a_{3}=0.66x10^{-6} m$. The averaging was performed in 100 simulations. In each case the volume fraction was fixed in 10\%, and the networks: Particle size $a_{1}=1x10^{-6}$ m, $Me=100x100$, $Mi=32x32$. Particle size $a_{2}=0.8x10^{-6}$ m , $Me=125x125 Mi=40x40 $. Particle size $a_{3}=0.66x10^{-7}$ m,  $Mi=47x47, Me=150x150$ }
\end{figure}

Simulations of $\lambda$ versus time for different particles sizes are presented in Figure 6. Theoretical curves were calculated using the values of the Einstein work for water at $T=290 K°$ . in 500 time steps and a volume fraction of approximately 10\%. Figure 6 also shows the comparison between theoretical and simulation values using previously particle sizes: $a_{1}, a_{2}, a_{3}$ and theoretical and computational closer results in 3 cases.

\begin{figure}
\includegraphics[scale=0.57]{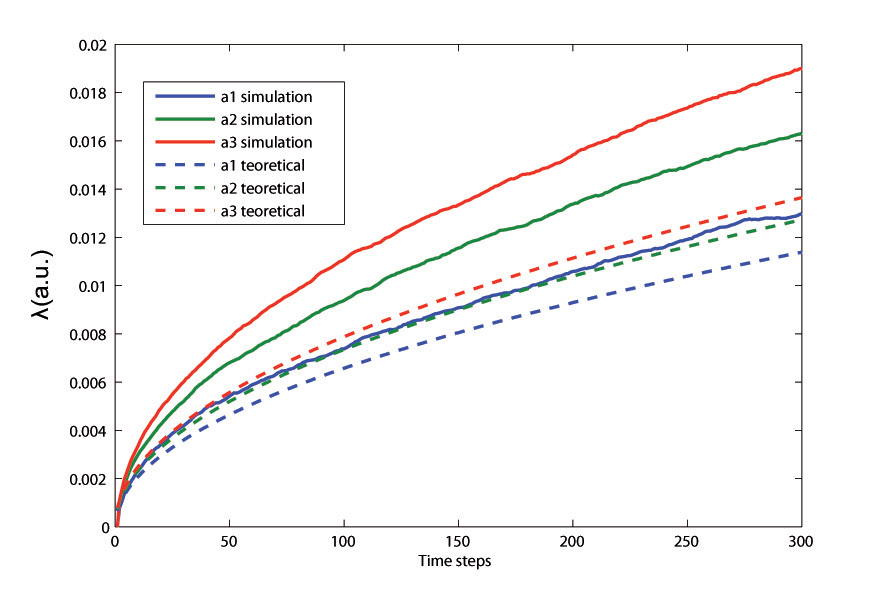}
\caption{$\lambda$ values as a function of time for particle sizes. $a_{1}, a_{2}, a_{3},$ and 500 time steps. Bridged lines represent the theoretical values and solid lines the experimental.  }
\end{figure}

Additionally, we have calculated the particles translational velocity as a function of time, in three scenarios: The first one, using particle sizes $a_{1}$ and $a_{2}$ and a volume fraction of approximately 10\% with a weight probability of movement of 20\% in each direction. In the second case, values were calculated for the particle sizes $a_{1}$ and $a_{2}$  in a volume fraction of 10\%, but varying the probability weights of the directions of motion as: 35\% down, 20\% no movement, 15\% up, 15 \% right, and 15\% left. For the third case we took the same kind of particles a1 and a2 but settings the weight probabilistic direction in: 50\% below, 20\% no movement, 10\% up, 10\% right and 10\% left.

The evolution of a1 and a2  are showed in figure 7. The case 1 shows the system in thermal equilibrium, where the particle velocities are constant in time\cite{08}. For case 2,  the particles tend to occupy lower positions as time is increased to be confined to the bottom of the simulation area, showing the evolution of the system to equilibrium, where velocities of particles decrease as time progresses. The case 3 is quite similar to case 2, except for a faster equilibrium transition because is assumed a bigger probability of movement in the down direction.

\begin{figure}
\includegraphics[scale=0.65]{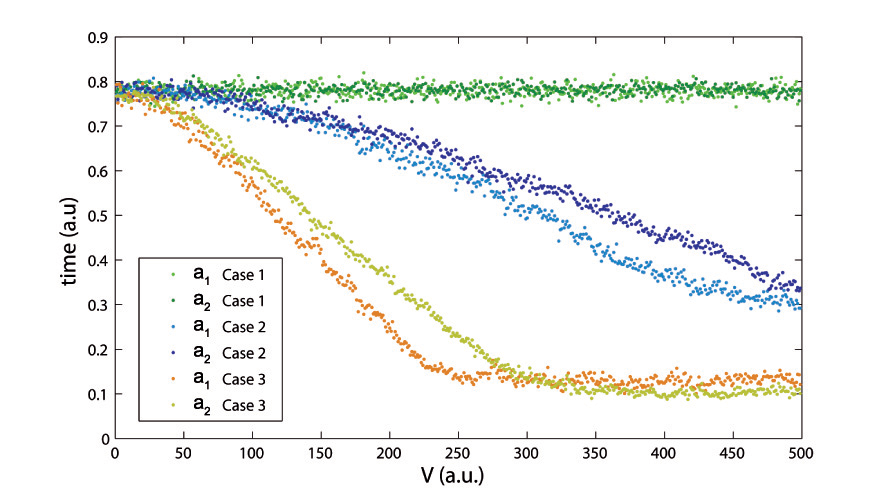}
\caption{ Translational velocity computations for three cases, with probabilities of movement: a) Case1 20\% down, up, right, left, and no movement. b) Case 2: down 35\%, 20\% no movement above 15\%, derecha15\%, 15\% left. c) Case 3: downward 50\%, 20\%  upward 10\%, right 10\%, 10\% left. }
\end{figure}

Finally the duration of simulations in three volume concentration values $(10\%, 25\% and 50\%)$ for matrixes of Me $(300x300=90000 cells, 250x250=62500 cells, 200x200=40000 cells, 100x100=10000cells and 50x50=2500 cells)$ is shown, for 100 time steps and using a movement probability of 20\% for each direction in Figure 7. The simulations were carried out in Matlab on a personal computer with Intel Core 2 Duo processor and 3GB of RAM. The results showed that, the heaviest calculus for 45000 particles, with a concentration of 50\% is performed in a little more of a couple of hours without paralleling neither algorithm optimization. A remarked increase of the computation time when the concentration arise is observed due a large numbers of operations that the programs make. Since the number of particles is not too bigger, the program allow to chance the parameters of simulations (the directions of particles movement) without adding more operations to the computing, because this changes only affect the probability direction, allowing to establish more complex situation without sacrifices the numbers of particles.

\begin{figure}
\includegraphics[scale=0.53]{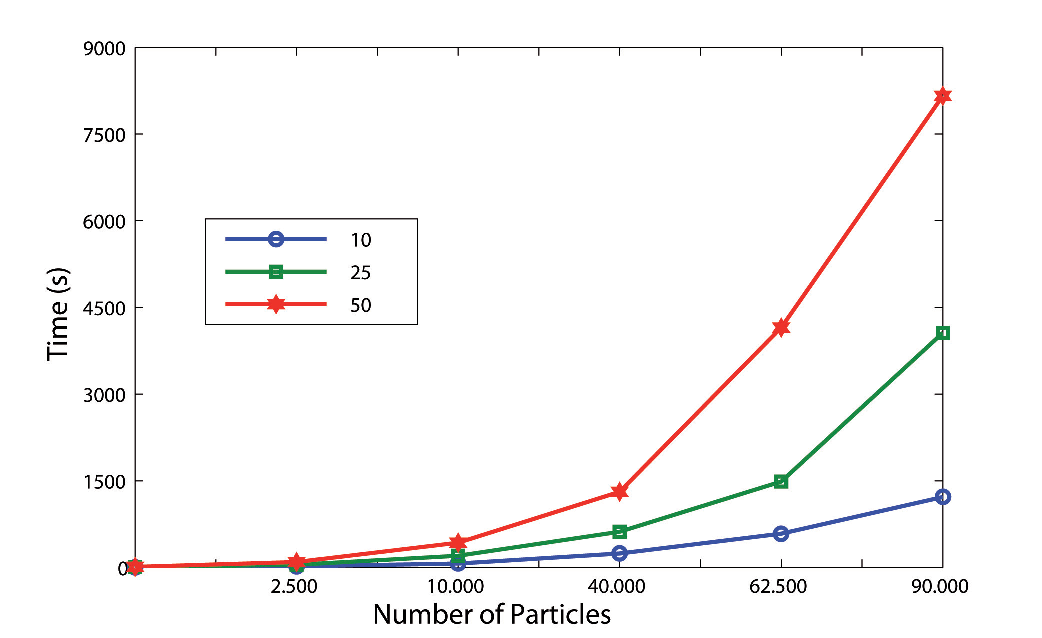}
\caption{  Simulations time duration for different values of concentration and different concentrations of particles. }
\end{figure}

\section{Conclusions}

The dynamics of Brownian particles was well presented by Reticular Mapping Matrix Method, validating results with typical mean square displacement of particles and relating this value to fluid self-diffusion coefficient, the result shown a good agreement between theoretical and simulation values. The calculus of translational velocity has shown the possibility to use this method in the thermal systems  equilibrium analysis, by controlling the particles probability movement direction. The operations of addition and subtraction of integers used by the simulation method allows a significant saving of computation time shown by a simulation of 45000 particles in two hours without parallelizing or optimization. We expect to use the scheme to allow the establishment of more complex simulation conditions that will be extending the applications of the algorithm.

\end{document}